# Efficient Similarity Indexing and Searching in High Dimensions

Yu Zhong

BAE Systems
6 New England Executive Park
Burlington, MA 01803-5012 USA
yu.zhong@baesystems.com

## Abstract

*Efficient indexing and searching of high dimensional data has been an area of active research due to the growing exploitation of high dimensional data and the vulnerability of traditional search methods to the "curse of dimensionality". This paper presents a new approach for fast and effective searching and indexing of high dimensional features using random partitions of the feature space. Experiments on both handwritten digits and 3D shape descriptors have shown the proposed algorithm to be highly effective and efficient in indexing and searching real data sets of several hundred dimensions. We also compare its performance to that of the state-of-the-art locality sensitive hashing algorithm.*

## 1. Introduction

Similarity indexing refers to the fast association of similar instances in a database when given a query. Search for nearest neighbors can be greatly expedited by only examining the associated items. With the rapid growth of the amount of multimedia data in today's information age, efficient similarity indexing has become more and more important in many fields including computer vision, pattern recognition, data mining, and machine learning.

Accompanying the explosion of information is the rising challenge of problems to be solved. Higher dimensional feature vectors have been used to tackle problems of increasing complexity. On the other hand, advances in computer hardware technology have made it possible for the storage and processing of a large amount of high dimensional data. Just a few years back, high dimensional spaces were usually referred as feature spaces of up to several dozen dimensions, mostly in texture analysis where features of different spatial and temporal characteristics were extracted, and in applications using eigen analysis. It is now common to work with feature spaces with a dimension of several hundred. For example, the bag of words model, widely used in natural language processing and information retrieval, represents documents using the frequencies of individual words in a dictionary. It was later adopted by computer vision researchers to represent complex objects for image classification and object recognition [38]. In recent years, invariant 2D image descriptors such as SIFT features [26] and 3D shape descriptors such as Spin Images [23], which are usually computed as local histograms of image or shape attributes, have gained increasing popularity due to their robustness, effectiveness, and ease of use for object recognition tasks. These local invariant descriptors have also been extended to audio analysis ("audio finger printing" [17]) as well as video analysis ("spatial-temporal event") to recognize complex audio and video events. In all these approaches, features of high dimensionalities are extracted to meet the challenges of highly complex problems. Even with recent advancements in learning and classification techniques including SVM and adaBoost, it has been argued that traditional nearest neighbor technical still remain to be competitive despite its ease of use in both training and querying [7].

Although sophisticated indexing algorithms [3][19] that work fairly well in low dimensional space have been developed, their performances deteriorate rapidly as the data dimension increases due to the curse of dimensionality: the complexity of the algorithm may grow exponentially in the dimensionality of the feature space. Efficient indexing of high dimensional features remains an active research area.

In this paper we present a simple but highly effective algorithm to search high dimensional feature spaces. We compute for a query a number of random convex hyper polyhedral neighborhoods in which each contains a controlled number of database points. Efficient indexing is achieved by only searching the set of database points in these neighborhoods. Although such point sets only consist of a very small fraction of all database points, these points are highly likely to be close enough to the query for the set to contain most, if not all, of its closest neighbors.

The remainder of this paper is organized as follows: Section 2 reviews published work on similarity searching and indexing in high dimensional spaces. Section 3 describes the new algorithm. Performance evaluations of the algorithm on two high-dimensional datasets are reported in Section 4. Section 5 concludes the paper with some discussions on the algorithm.

## 2. Previous work

Many of the published indexing algorithms for high dimensional space either a) first reduce the original space to a lower dimensional space where standard indexing is sufficient [20] or b) use variants of standard indexing trees [5][11][33][40] to alleviate performance degradation when the data dimension is moderately high. A comprehensive survey of these methods can be found in [6][10]. Traditionally, index trees have used "branch and bound" or "backtrack" techniques to prune tree nodes and exclude a large portion of database data from being searched to achieve efficient indexing. However, these techniques become ineffective in high dimensional search, since the space enclosed by a node almost always intersects the search volume as dimension grows. A variant of the K-d tree search algorithm was used in [2] to index features for shape matching where moderate performance gain was achieved by examining tree nodes in order of increasing distance from the query to find approximate nearest neighbors. Experiments showed improvements in feature spaces of 5 to 25 dimensional spaces, with computational cost picking up toward the high end. Despite the improvements in indexing data of moderately high dimensions, it has been suggested [4][39] that using the multi-dimensional indexing structures for searching becomes less efficient than a sequential scan when the data dimension increases beyond a certain size.

Nene and Nayar [31] proposed a simple algorithm to find the nearest neighbor within a user specified distance *r*. Their algorithm works by quickly trimming database points that are at a distance *r* away from the query in each of the *d* coordinates using a data structure that facilitates fast look-up between the set of raw data points and *d* ordered set sorted using each coordinate. However, their approach is less suited for K-NN search where we may not know the range of distances between nearest neighbors.

Randomized algorithms have been proposed to perform fast approximate nearest neighbor search in high dimensional spaces [16][22][27]. Locality sensitive hashing (LSH)[16][21][22], named after the fact that random binary tests on subspace projections of high dimensional features were used to generate hashing functions preserving the distance between points statistically, has been practically used to successfully index data with dimensions exceeding one hundred [13][14][18][28][29][32][34]. The LSH approach constructs a hash table of $2^K$ entries using *K* random binary tests on the features. The hash functions are locality sensitive in that they are likely to map similar features into the same bucket. A number of *L* such random hash tables are generated. At query time, a query is hashed to a bucket in each of the hash tables. The query is only compared to the features in those buckets, which usually consists of a small fraction of the original database in order to gain huge reduction in computation.

As popular as LSH is, we note some weaknesses of the approach, mostly due to the rigidity common to all hash tables. First, it is difficult to determine the resolution for the hash table. If the resolution is too fine, many buckets become empty, which increases the risk of a "miss" where a query is hashed to an empty bucket and no match is found. If the resolution is too coarse, then often there are too many points in the hashed bucket which need to be searched, resulting in an increase in search cost. The LSH is ideal for the *r*-NN problem which computes all neighbors within a pre-specified distance *r* to the query, but sub-optimal for the more common *k*-NN problem which computes the *k* nearest neighbors to a query where the nearest distance may vary from point to point, a scenario that is very typical in real applications. It is difficult for LSH to compromise between the need to keep a moderate number of points in a bucket to reduce the amount of search and at the same time to ensure enough points are in a bucket to contain a good nearest neighbor with a LSH of a fixed resolution when the data density varies. It has been suggested that such problems be approached using a cascade of LSH tables constructed at multiple resolution levels [16][22][27]: the LSHs are searched in order of decreasing resolution, until either a match is found or all hash tables have been searched. Although the use of multiple LSHs improves the search efficiency, it is as good as an approximation can get, at the cost of extra computational resources. Secondly, for LSH to be effective, the number *K* of hash keys is suggested to be a certain proportion (between 10% and 30%) of the data dimension. As a result the number of buckets grows exponentially with dimension *d*. For feature spaces with dimension beyond a couple hundred, the hash table becomes too large to be loaded in memory or disk. LSH works around this by introducing a secondary traditional hash table to hash only the non-empty buckets from the primary locality sensitive hashing. This secondary hashing is not locality sensitive and may reduce the overall performance.

Randomized forest [1][8][9][12][15][25][36][37] is a branch of machine learning methods which constructs an ensemble of independent random tree classifiers for classification and recognition applications. Each internal node of the trees contains a random test on the features to dispatch an incoming data item to one of the child nodes. Each leaf contains an estimate of the conditional distribution over the classes based on the training data points at the node. A query point is dropped down each of the trees using the stored test at each node it encounters and then classified based on the aggregated distribution estimate at the arrived leaf nodes from all trees. Randomized KD trees [37] have been used to compute approximate nearest neighbors in high dimensional space. Such approaches construct multiple KD trees and search them simultaneously until a pre-specified number of leaf nodes have been examined to limit the amount of backtracking. A priority queue for the nodes to be searched is usually maintained and shared among the trees to speed up the search. Although significant improvements have been achieved, it still uses backtrack which is ineffective in searching high dimensional space.

This paper presents a randomized algorithm for fast similarity search in high dimensional spaces. We use random partition trees to obtain disjoint and complete data adaptive partitions of the feature space consisting of convex hyper polyhedrons among which database points are evenly distributed. A fast mapping associates a query point with a polyhedral cell where all its enclosed database points are retrieved. Multiple trees are used to guard against quantization errors. In contrast to randomized KD trees, neither backtrack nor branch and bound is used in the search. As a result, there is no need for a priority queue and the search of each tree can be performed independently. It also requires minimum indexing overhead to retrieve a small set of points for distance comparison: only one random coordinate access of the query and one float number comparison for each node visited. The proposed algorithm also compared favorably to LSH as the partitions adapt to the data distribution in that data points are relatively evenly distributed in the bins. We describe the new algorithm in detail in the following section.

## 3. Algorithm

In this paper, we adopt the convention to categorize a $d$-dimensional space in the context of the size $N$ of the data set: the feature space is considered to be high dimensional if $N << 2^d$, low dimensional if $N >> 2^d$, and moderate otherwise.

The underlying idea of the algorithm is to search random convex hyper polyhedral neighborhoods of a query for its closest neighbors. In particular, each of the random neighborhoods is deliberately constructed to contain a controlled number of database points – neither too many, nor too few, for indexing accuracy and efficiency. We use an easily computed partitioning technique to randomly divide the high dimensional feature space into non-overlapping cells of convex hyper polyhedrons bounded by hyper-planes. Furthermore, the partition adapts to the data distribution in that database points are relatively evenly distributed among the bins. A number of such random partitions are computed independently. Each renders a fast mapping from an arbitrary query point to a convex hyper polyhedron that encloses both the query and a non-empty subset of database points. Although the union of all these random convex polyhedral neighborhoods of the query may only contain a tiny fraction of the database points, they are highly likely to enclose all the nearest neighbors to the query. Drastic reduction in computation is achieved by only searching these points in the similarity analysis.

We propose to use randomized forest to partition the high dimensional feature space into non-overlapping convex hyper-polyhedral cells which relatively evenly divide the dataset points. A randomized binary partition tree is constructed such that each node of the tree corresponds to a convex hyper-polyhedron containing a nonempty subset of data points, with the root being the entire feature space. We use the random test at each internal node to define a hyper-plane subdividing its hyper-polyhedron into two hyper-polyhedrons -- one for each child node. All the leaf nodes form a non-overlapping and complete partition of the feature space, with each cell containing a controlled number of database points. A number of such random partitions are computed for a database using randomized forest. During search, a random convex hyper-polyhedral neighborhood of a novel query is computed as the arrived leaf polyhedron by dropping the query down each pre-computed partition tree. The database points in these polyhedral neighborhoods of the query are then searched for its near neighbors.

We use the following notations: The database $P = \{p_0, p_1, p_2, \cdots, p_{N-1}\}$ consists of $N$ data points in $d$-dimensional space where $p_i = (p_{i0} \quad p_{i1} \quad \cdots \quad p_{id-2} \quad p_{id-1})^t$. We define a random binary partition forest $F$ of size $L$, split ratio $r$ where $(0 < r \leq 0.5)$, and capacity $C$ as a set of $L$ random binary partition trees $F = \{T_0, T_1, \cdots, T_{L-1}\}$, where the $j$-th internal node of tree $T^i$ contains a binary test $t^i_j$, and each leaf node contains between $r \bullet C$ to $C$ database points.

### 3.1. Random tests

We design a simple random test at each internal tree node to assign an incoming data point to one of its two children. The test is based on subspace projections of high dimensional features. Suppose the node $l_i$ contains $n_i$ data points $x_0, x_1, \cdots, x_{n_i-1}$ where $x_j = (x_{j0} \quad x_{j1} \quad \cdots \quad x_{jd-1})^t$. We randomly select an index set $I_i = \{d_{i0}, d_{i1}, \cdots, d_{ik-1}\}$ of size $K$ ($K \le d$) where $d_{ij} \in (0,1,\cdots,d-1)$ and random coefficients $\mathfrak{I}_i = \{\xi_{i0}, \xi_{i1}, \cdots, \xi_{ik-1}\}$ where $\xi_{ij} \in [0,1]$, and compute for every feature vector $x_j$, a scale value $y_j = \sum_{k=0}^{K-1} x_{jd_{ik}} \xi_{ik}$ for $j = 0, \cdots, n_i - 1$. We then sort the $y_j$s and denote the sorted sequence as $\{\tilde{y}_j\}$. Randomly select $\psi_i$ between the $100 \cdot r$ and $100 \cdot (1-r)$ percentile points of $\{\tilde{y}_j\}$, ie., $\psi_i \in [\tilde{y}_{n_i \cdot r}, \tilde{y}_{n_i \cdot (1-r)}]$. The random test

$$t_i(x) = \sum_{k=0}^{K-1} x_{d_{ik}} \xi_{ik} - \psi_i \ge 0 \qquad \text{Eq. 1}$$

defines a hyper-plane which splits the polyhedron at node $l_i$ into two polyhedrons each containing no less than $r \cdot n_i$ data points.

### 3.2. Algorithm to build the random binary partition forest

Now we describe an algorithm to build the indexer of a random partition forest incrementally by adding one data point at a time. Starting with an empty root/leaf node, we proceed as follows to build a random binary partition tree:

– Randomly select without replacement a database data point $p_i$;
– Drop $p_i$ down the tree until it reaches a leaf node based on the tests at the internal nodes it encounters;
– If the number of database points at the leaf node exceeds $C$, split the node into two children using a random hyper-plane generated according to Eq. 1. Save the random test with the now internal node.
– Repeat the above steps until all data points have been inserted in the tree.

An ensemble of $L$ such trees are generated to form $L$ random partitions of the feature space. The trees, including random tests at the internal nodes and indices of database points at the leaf nodes, are saved for efficient search and query of the database. **Figure 1** presents the pseudo code for this algorithm.

### 3.3. Query and Search

As described in the previous section, we build a pyramid of embedding convex hyper-polyhedral partitions of the high dimensional feature space using the randomized binary partition tree where each leaf node corresponds to a non-overlapping convex volume containing a controlled number of database points. Each randomized binary partition tree facilitates a fast mapping from an arbitrary data point in the high dimensional space to the random hyper-polyhedral neighborhood enclosing that point, or equivalently, a controlled number of database points that are contained in the same convex polyhedron. During query time, a query data point descends each of the pre-computed randomized trees according to the outcome of the binary test at each internal node it encounters until it reaches a leaf node. The database points stored at these leaf nodes are retrieved for comparison with the query. The pseudo code for this process is shown in Figure 3. **Figure 2** illustrates the query process. Although this retrieved data set usually only consists of a tiny fraction of the original database, it is highly likely to contain most, if not all of the database points close to the query. Highly accurate and efficient search of high dimensional dataset is achieved this way.

```
TrainTrees(TreeNode T[], DataPoint X[])
{
  for(i=0; i<L;i++)
    TrainOneTree(T[i], X);
}

TrainOneTree(TreeNode T, DataPoint X[])
{
  randomly_for_each(i in 0:N-1)
  {
    TreeNode node = T;
    while (!node.IsLeaf())
    {
      if(node. randomTest (X[i]) == true)
        node = node.leftChild;
      else
        node = node.rightChild;
    }
    node.addPoint(X[i]);
    if(node.GetNumerOfDataPoint() > C)
    {
      RandomTest t = RandomTest (node.GetDataPoints(), r);
      node.randomTest = t;
      TreeNode  left(t.GetPassDataPoints(node.GetDataPoints()));
      TreeNode  right(t.GetFailDataPoints(node.GetDataPoints()));
      node.leftChild = left;
      node.rightChild = right;
      node.ClearDataPoints();
    }
  }
}
```

Figure 1. Pseudo code for constructing the random binary partition forest in sequential mode.

### 3.4. Discussions

There are four parameters in the algorithm, namely, $L$, the number of random partitions used, $r$, the split ratio, and capacity $C$, the maximum number of database points in a leaf node, and $K$, the number of indices used in the subspace projection in the random tests.

The capacity $C$ determines the expected depth of the partition trees, or equivalently, the resolution of the partitioning cells. As depth increases, the leaf cells are divided further and the similarity among points in a leaf cell likely increases. A smaller $C$ results in increased retrieval precision.

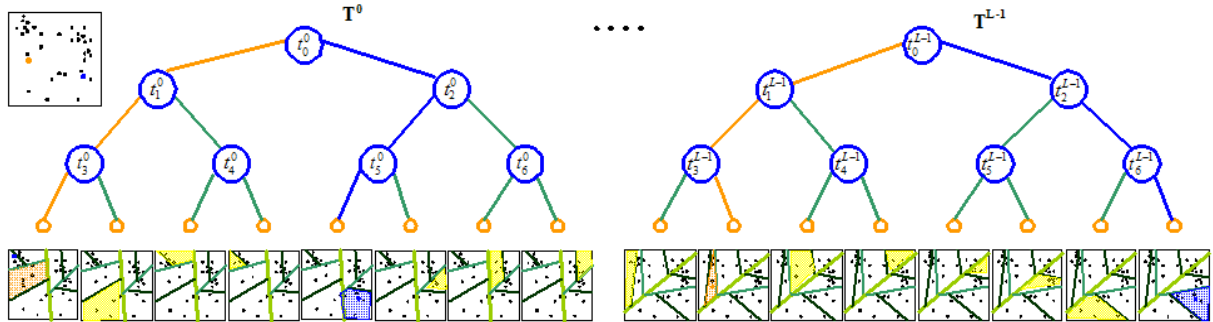

Figure 2 Query the random binary partition forest. Feature space is shown in the upper left corner with the query points marked in color and database points marked in black. The path that a query descends the random binary partition tree is highlighted in the same color as the query. The mapped convex cell for a query is also highlighted in the same color. Only database points in these cells are searched in similarity analysis.

```
DataPoints Retrieve (TreeNode  T[], DataPoint x)
{
   DataPoints retrievedPoints = ∅ ;
   For(i=0; i<L;i++)
      retrievedPoints = retrievedPoints ∪ Retrieve (T[i], x);
   return retrievedPoints;
}

DataPoints Retrieve(TreeNode  T, DataPoint  x)
{
   TreeNode  node = T;
   While (!node.IsLeaf())
   {
      if(node. randomTest (x) == true)
         node = node.leftChild;
      else
         node = node.rightChild;
   }
   return node.GetDataPoints();
}
```

Figure 3 Pseudo code for retrieving a small set of database point for a query using the pre-computed random partitions.

The split ratio $r$ controls variations in the number of database points in the polyhedral cells. Indirectly, it affects the variation in the depth of the leaves. A $r$ close to *0.5* likely creates more balanced trees.

As each leaf cell contains between $r \cdot C$ to $C$ database points, the partition of the feature space has a varying resolution that adapts to the data distribution: more cells are likely generated where data points are dense while fewer cells are created where data points are sparse. This helps to improve the indexing efficiency of the algorithm.

The use of only one random partition is likely to result in mistakes as points nearby to a query may be missed depending on the actual hyper-planes defining the partition and the locations of the near points. The close points missed by one random hyper-polyhedral neighborhood are likely enclosed by other independently generated random hyper-polyhedral neighborhoods. A large $L$ helps to reduce the miss rate and improve the recall. If the accuracy using one tree is $p$, then under the assumption of independency of the trees, we can boost the accuracy to $1-(1-p)^L$ by combining the results from the $L$ trees.

The number of indices $K$ used in subspace projection determines how many data coordinates are involved in the splitting hyper-planes. A larger $K$ in general provides better randomization. On the other hand, considering each internal node as a binary classifier, $K$ is the number of features, good or bad, averaged to provide a new feature fed to the classifier. A large $K$ means more features are averaged which may produce a less discriminative feature. Our empirical study indicates that the indexing performance improves slightly as $K$ increases from 1 to 2 or 3, and than starts degrading as $K$ increases further, in a manner similar to the effect of $K$ in $K$-nearest neighbor classification. In this paper, we focus on the simple case when *K=1*. In this case, the split function randomly selects a coordinate, sorts the data points using the values at the coordinate, and splits them at a random threshold. This results in a very simplified algorithm: each internal node contains one coordinate indice and one threshold to define a random axis parallel hyper-plane to separate data points into two sets. Only comparisons and random number generations are involved during the construction of the indexer of random partitions, and an average of $L \cdot \log_2^{2N/(1+r)C}$ float number comparisons are all that's required to retrieve the indexed data point set for a query.

We note that both the construction of the indexing structure and the search using the proposed algorithm involves little optimization. The expected complexity to compute the random partitions is $O(L \cdot N \cdot \log(N))$. The storage for the random partitions is $O(L \cdot N)$. The complexity of the expected indexing time to retrieve the data points for comparison is $O(L \cdot \log(N))$.

## 4. Experiments

The proposed algorithm has demonstrated both high accuracy and efficiency in searching datasets of several hundred dimensions from real applications. It has also been shown to outperform the state of the art LSH algorithm. All the experiments are run on a 2.4GHz processor with a maximum of 2GB memory for each user process.

One bench mark dataset for high dimensional search is the MNIST database of images of handwritten digits [24]. The database contains *60,000* examples, and the test set has *10,000* examples. A label from 0 to 9 is assigned to each database and training image. Each image is of size $28 \times 28$ which translates into a vector of *784* dimensions using the raw image array.

We have applied the proposed algorithm to search the MNIST database. Each feature vector is first normalized to have a norm of 1. We build the random partitions using the *60,000* examples in the database, which is then used to retrieve a small set of database points for each of the examples in the test dataset. The nearest neighbor in the retrieved set is computed for each query using the Euclidian distance and compared to the exact nearest neighbor. We use the percentage of correctly computed nearest neighbors as the accuracy measurement.

In the experiment, *K* is set to one so that only one random index is used and the splitting hyper-planes become axis parallel and the partition cells degenerate to hyper cuboids. The split ration *r* is set to 0.3. The capacity of each tree node is set to *12* which results in a depth of around *13* for the partition trees for the database of *60,000* features. Therefore, all it takes to retrieve the indexed set for a query is *13*L* float number comparisons. We set the number of trees *L* to be *1, 2, 5, 10, 20, 40, 80, 160, 320*, and *640*. For each parameter set, the percentage of correctly computed nearest neighbors using the drastically reduced indexed set is used to measure the search accuracy. The average size of the indexed set, ie., the number of database points searched for each query is used to indicate the search cost. A total of *20* random trials are performed for each parameter setting. The average search accuracy versus the search cost is shown in **Figure 4**. While using one random partition gives *7.7%* search accuracy by only examining less than *9* out of the *60,000* database points to compute the nearest neighbor for a query, the accuracy quickly improves to *96.1%* using *80* trees which reduces the dataset examined to only *0.9%* of its original size. A near perfect accuracy of *99.99%* is achieved using *640* random partitions, with *4.7%* of the database points searched.

We have run the LSH algorithm (http://www.mit.edu/~andoni/LSH) on the same dataset. A cascade of four LSHs of radii *0.4, 0.53, 0.63*, and *0.88* is used. The max radius of *0.88* is set so that the nearest neighbors to all data points in the dataset are within this value. We specify a variety of accuracy parameters valued between *0.04* and *0.99* as input to the software and let it compute the optimal LSH tables for the dataset with the desired accuracy and a memory constraint of *2.0* GB. **Figure 4** shows the performance of the multi-layer LSH on the dataset. The proposed algorithm shows significantly superior performance, searching just a small fraction of database points to achieve the same nearest neighbor accuracy comparing to LSH.

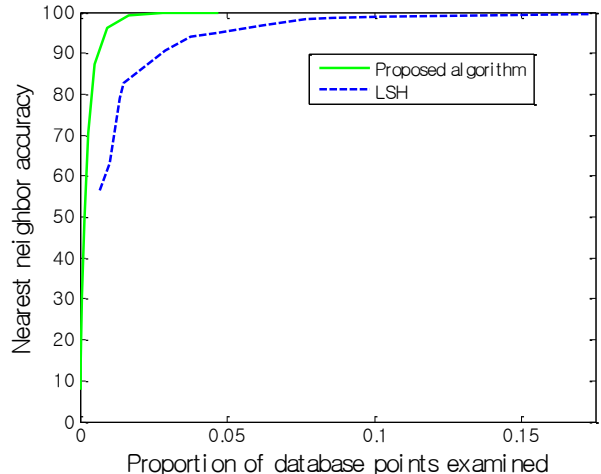

Figure 4. Nearest neighbor search performance comparison between the proposed search algorithm and the LSH approach on the 784D MNIST dataset for hand written digit classification.

We also evaluate the algorithm on a high dimensional dataset for 3D object recognition. The feature set is computed using the Intrinsic Shape Signature (ISS) by Zhong [41], a view invariant 3D shape descriptor for 3D object recognition. ISS represents the local shape patch using a weighted point occupational histogram of data points in a support volume w.r.t. a view invariant local coordinate system. This invariant representation of local 3D shape is high dimensional to be discriminative of subtle shape differences. We extract the ISS shape descriptors for point clouds from *72* 3D vehicle models from the Princeton Shape Benchmark [35] to form a database of *250,736* feature vectors of *595* dimensions. Two partial views of each model are simulated and *30,000* test features are extracted from the test point clouds of partial views to form the test set. We search the database to compute the nearest neighbor for each test feature using the proposed algorithm. Instead of using the Euclidean distance, we used the Chi-Square divergence as suggested in [41] $dist(x_i, x_j) = \sum_{k=0, d-1} (x_{ik} - x_{jk})^2 / (x_{ik} + x_{jk})$ as the distance measure. The capacity of the leaf nodes is set to be *12*, which results in an average tree depth of *15* for the database of *250,736* features. We evaluate indexers of an increasing number of random partitions and plot the percentage of computed true nearest neighbors versus the portion of database features examined as shown in **Figure 5**. With 40 random partitions, more than *69%* of exact nearest neighbors are retrieved with less than *0.13%* of database points examined. The accuracy improves to *91%* and *96%* with *0.48%* and *0.91%* of database points searched using *160* and *320* trees, respectively. While a nearest neighbor query of the database of *250,736* 595D features using exhaustive search takes an average of 0.73 sec, the proposed algorithm reduces the average query time to 0.009 sec with an accuracy exceeding 96% on the test set, which is a 81x speedup including all the indexing overhead. We also shown in **Figure 5** the performance using LSH, obtained by specifying a success probability of 0.95 at various values for the radius R in R-NN search, using a maximum memory of 2GB. While the LSH performance degrades notably when computing nearest neighbors for more query points, the proposed algorithm performs much better thanks to its adaptive nature to the data distribution.

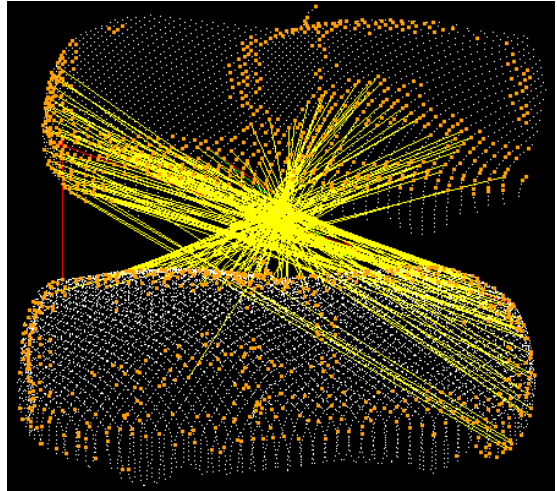

Figure 6. Matches between shape descriptors of a car model (bottom) and a partial view query of the car (top). Shape descriptors are extracted at key points (in orange) from the two 3D point clouds shown in white. Yellow lines show matches where both the ENN and the proposed algorithm agree. The matches computed by only the ENN, or only the proposed algorithm are drawn in red and green respectively.

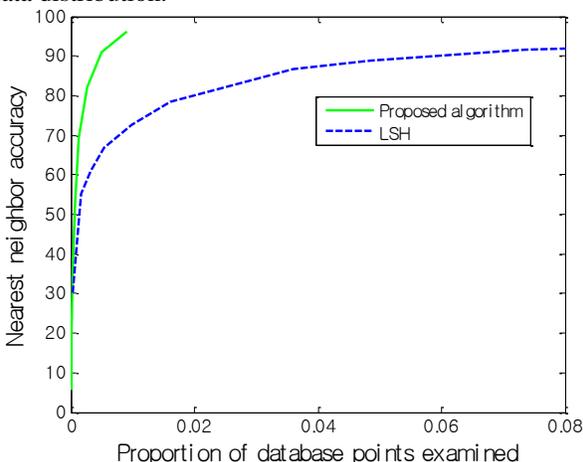

Figure 5. Nearest neighbor search performance comparison between the proposed algorithm and the LSH approach on the dataset of 595D shape features for 3D object recognition.

We show in Figure 7 the matches established between local shape volumes from a car model (bottom) in the database and a query point cloud (top) using 40 trees with a capacity of 12 and compare them to the exact nearest neighbor results. One-to-one pair-wise matches are established using a greedy algorithm. The matches with distances less than a threshold are shown where yellow lines correspond to matches where both the ENN and the proposed algorithm agree. The matches computed by only the ENN, or only the proposed algorithm are drawn in red and green respectively Although the indexing algorithm searches only *0.13%* of the database features, it retrieves very similar results as the ENN algorithm for these matches, with only *4* discrepancies among the *138* matches.

## 5. Conclusions

This paper presents a highly effective and efficient similarity searching algorithm for high dimensional space. It has many ideal properties such as ease of training and sub-linear time for search and indexing. The algorithm makes use of redundant sampling of random neighborhoods and requires little optimization. The fact that each random partition or tree is independently constructed and searched makes the algorithm easily parallelizable and distributable. It also supports incremental updating of the database at little computational cost – upon the query of a new data point, we can easily update the indexer by saving the novel point in the arrived leaf node and split the node when necessary. The proposed algorithm also appears to effectively compute similarity association, with different distance measures that are application dependant.

## References

[1] Y. Amit and D. Geman. Shape quantization and recognition with randomized trees, Neural Computation, 9:1,545-1,587.
[2] J. Beis and D. G. Lowe, Shape indexing using approximate nearest-neighbour search in high-dimensional spaces. Proc. CVPR, pp. 1000-1006, 1997.
[3] J. Bentley, Multidimensional binary search trees used for associative searching. Commun. ACM 18(9):509–517.
[4] S Berchtold, C Böhm, D Keim, H Kriegel, A cost model for nearest neighbor search in high-dimensional data space, Proc. 16th ACM PODS, pp. 78-86, 1997.

[5] S. Berchtold, D. Keim, and H. Kriegel The X-tree: An Index Structure for High-Dimensional Data, - Readings in Multimedia Computing and Networking, 2001.

[6] C. Bohm, S. Berchtold, and D. Keim. Searching in high-dimensional spaces: Index structures for improving the performance of multimedia databases. ACM Computing Surveys, 33(3):322-373, 2001.

[7] O. Boiman, E. Shechtman and M. Irani. In defense of Nearest Neighbor based image classification, Proc. CVPR, pp: 1-8, June 2008.

[8] A. Bosch, A. Zisserman and X. Munoz. Image Classification using Random Forests and Ferns. Proc. ICCV 2007.

[9] L. Breiman, Random Forests, Machine Learning, 45(1):5-32, 2001.

[10] E. Chavez, G. Navarro, R. Baeza-Yates, and J. Luis Marroquin. Searching in metric spaces. ACM Computing Surveys, 33(3):273:321, 2001.

[11] P. Ciaccia and M. Patella: PAC Nearest Neighbor Queries: Approximate and Controlled Search in High-Dimensional and Metric Spaces. Proc. 16th Int'l Conf. on Data Engineering, pp. 244-255, 2000.

[12] A. Criminisi, J. Shotton, and E. Konukoglu. Decision forests for classification, regression, density estimation, manifold learning and semi-supervised learning. *Microsoft Research, Tech. Rep. MSRTR-2011-114* 5.6 (2011): 12.

[13] T. Dean, J. Yagnik, M. Ruzon, M. Segal, J. Shlens, and S. Vijayanarasimhan Fast, Accurate Detection of 100,000 Object Classes on a Single Machine, Proc. CVPR 2013.

[14] B. Georgescu, I. Shimshoni and P. Meer. Mean shift based clustering in high dimensions: A texture classification example, Proc. ICCV, pp. 456-463, 2003.

[15] P. Geurts, D. Ernst and L. Wehenkel. Extremely randomized trees, Machine Learning, 36(1):3-42, 2006.

[16] A Gionis, P Indyk and R Motwani, Similarity Search in High Dimensions via Hashing, Proc. of the 25th Int'l Conf. on Very Large Data Bases, pp. 518-529, 1999.

[17] J. Goldstein, J. Platt and C. Burges, Redundant Bit Vectors for quickly searching high-dimensional regions, Deterministic and statistical methods in machine learning, Springer lecture notes on Computer Science 3635, pp. 137-158.

[18] K. Grauman and T Darrell, Pyramid Match Hashing: Sub-Linear Time Indexing Over Partial Correspondences, Proc. CVPR, 2007.

[19] A. Guttman, R-trees: a dynamic index structure for spatial searching. Source. Readings in database systems, pp. 599 – 609, 1988.

[20] J. Hafner, H. Sawhney, W. Equitz and M. Flickner, Efficient Color Histogram Indexing for Quadratic Form Distance Functions, IEEE Trans PAMI, 17(729-736), 1995.

[21] P. Indyk. Nearest neighbors in high-dimensional spaces. Handbook of Discrete and Computational Geometry, chapter 39, CRC Press, 2nd edition, 2004.

[22] P. Indyk and R. Motwani, Approximate Nearest Neighbors: Toward removing the curse of dimensionality, 30th ACM symposium on Theory of computing, pp. 604-613, 1998.

[23] A. Johnson and M. Hebert. Using spin images for efficient object recognition in cluttered 3D scenes. *IEEE Trans PAMI*, 21(5):433-449, 1999.

[24] Y. Le Cunn. The mnist database of handwritten digits. http://yann.lecun.com/exdb/mnist.

[25] V. Lepetit, P. Lagger and P. Fua, Randomized trees for real-time keypoint recognition, Proc. CVPR, 2005.

[26] D. Lowe, "Distinctive Image Features from Scale-Invariant Keypoints". IJCV 60 (2): 91–110, 2004.

[27] E. Kushilevitz, R Ostrovsky and Y. Rabani, Efficient search for approximate nearest neighbor in high dimensional spaces. SIAM J. Comput., 30(2):457-474, 2000.

[28] B. Matei, Y. Shan, H.S. Sawhney, Y. Tan, R. Kumar, D. Huber, and M. Hebert. Rapid object indexing using locality sensitive hashing and joint 3D-signature space estimation, IEEE Trans. PAMI, (28)7:1111-1126, 2006.

[29] G. Mori, S. Belongie and H. Malik. Shape contexts enable efficient retrieval of similar shapes, Proc. CVPR, 2001.

[30] M. Muja and D. G. Lowe, "Fast approximate nearest neighbors with automatic algorithm configuration," Int'l Conf. on Computer Vision Theory and Applications, 2009.

[31] S. A. Nene and S. K. Nayar, "Closest Point Search in High Dimensions," Proc CVPR, pp.859-865, 1996.

[32] A. Qamra, Y. Meng and E. Chang. Enhanced Perceptual Distance Functions and Indexing for Image Replica Recognition, IEEE TRANS. PAMI, 27(3):379-391, 2005.

[33] Y. Sakurai, M. Yoshikawa, S. Uemura, and H. Kojima, The A-tree: An Index Structure for High-Dimensional Spaces Using Relative Approximation, Proc. 26th Int'l Conf. on Very Large Data Bases, pp. 516-526, 2000.

[34] G. Shakhnarovich, P. Viola, T. Darrell, Fast pose estimation with parameter-sensitive hashing, Proc. ICCV, 2003.

[35] P. Shilane, P. Min, M. Kazhdan and T. Funkhouser. The Princeton Shape Benchmark, Shape Modeling International, 2004.

[36] J. Shotton, M. Johnson and R. Cipolla. Semantic Texton Forests for Image Categorization and Segmentation, Proc. CVPR, 2008.

[37] C. Silpa-Anan and R. Hartley, Optimised KD-trees for fast image descriptor matching. Proc. CVPR pp. 1-8, 2008.

[38] J. Sivic and A. Zisserman, Efficient visual search of videos cast as text search, IEEE Trans PAMI, 31(4):591-606, 2009.

[39] R. Weber and P. Zezula. A quantitative analysis of performance study for similarity-search methods in high-dimensional spaces. Proc. 24th Int'l conf. on Very Large Databases, pp. 194-205, 1998.

[40] D. White, R. Jain, Similarity Indexing with the SS-tree, Proc. Int'l Conf on Data Engineering, pp.516-523, 1996.

[41] Y. Zhong, "Intrinsic Shape Signatures: A Shape Descriptor for 3D Object Recognition", ICCVW, 2009.

## Acknowledgement

We would like to thank Alex Andoni of MIT for kindly providing the LSH implementation and answering questions about the LSH algorithm. This research was developed with funding from the Defense Advanced Research Projects Agency (DARPA). The views, opinions, and/or findings contained in this article are those of the author and should not be interpreted as representing the official views or policies of the Department of Defense or the U.S. Government. Approved for public release; distribution is unlimited.